\documentstyle[12pt]{article}
\newlength{\abstractwidth}
\begin{document}
\thispagestyle{empty}
\pagestyle{plain}
\def\beq{\begin{eqnarray}}
\def\eeq{\end{eqnarray}}
\def\nn{\nonumber\\}
\renewcommand{\thefootnote}{\fnsymbol{footnote}}
\renewcommand{\thanks}[1]{\footnote{#1}} 
\newcommand{\starttext}{
\setcounter{footnote}{0}
\renewcommand{\thefootnote}{\arabic{footnote}}}
\begin{titlepage}
\bigskip
\hskip 3.7in\vbox{\baselineskip12pt
\hbox{CGPG-00/5-1}\hbox{SINP/TNP/00-11}}
\bigskip\bigskip\bigskip\bigskip

\begin{center}
{\large \bf STATISTICAL ENTROPY OF SCHWARZSCHILD\\[.4cm]
BLACK STRINGS AND BLACK HOLES}
\end{center}

\bigskip\bigskip
\bigskip\bigskip

\centerline{\bf Saurya Das$^a$\thanks{das@gravity.phys.psu.edu}, 
Amit Ghosh$^a$\thanks{ghosh@gravity.phys.psu.edu} and 
P. Mitra$^b$\thanks{mitra@tnp.saha.ernet.in}}
\medskip
\begin{center}
$^a$Center for Gravitational Physics and Geometry, Department of 
Physics,\\ The Pennsylvania State University,  
University Park, PA 16802-6300, USA\\
$^b$Saha Institute of Nuclear Physics, 1/AF Bidhannagar,
Calcutta 700064, INDIA
\end{center}

\bigskip\bigskip

\begin{abstract}
\baselineskip=16pt
The statistical entropy of a Schwarzschild black string in five
dimensions is obtained 
by counting the black string states which form a representation of 
the near-horizon conformal symmetry with a central charge. The 
statistical entropy of the string agrees with 
its Bekenstein-Hawking entropy as well as that 
of the Schwarzschild black hole in four dimensions. The 
choice of the string length which gives the Virasoro algebra 
also reproduces the precise value of the Bekenstein-Hawking  
entropy and lies inside the stability bound of the string. 
\end{abstract}
\vspace{3cm}
\begin{flushleft}
\end{flushleft}
\end{titlepage}
\starttext
\baselineskip=18pt
\setcounter{footnote}{0}


The aim of this paper is to statistically count all 
microstates which represent a five dimensional Schwarzschild 
black string of 
a given mass per unit length in the limit $\ell_P^{\,2}/R_s^2
\sim G/R_s^2\to 0$, where $R_s$ is the Schwarzschild radius. 
The microstates form a microcanonical 
ensemble for the classical Schwarzschild black string. The 
Bekenstein-Hawking (BH) entropy of the string is the same as 
that of a Schwarzschild black hole in a spacetime of one lower 
dimension (because $G_n=\ell G_{n-1}$). We show that
the Boltzmann entropy of the string agrees with its BH entropy
and argue how one may reproduce the Boltzmann entropy of the
black hole from that of the string. 

So far statistical computations of entropy have been limited 
to near-extremal black holes in one approach \cite{Strom} and
non-rotating black holes in another approach \cite{Abhay}.
There has been another attempt \cite{Carlip} which skips the 
details of the underlying quantum theory and relies more on near-horizon 
symmetries. Our analysis is in the same spirit. However, in
the detailed framework of \cite{Carlip}, an essential use
of the $(t-r_*)$-plane of black holes seems rather artificial.
Moreover, it is not directly applicable to the Schwarzschild 
case. Here we propose an alternative way of 
getting an infinite (conformal) symmetry in the near-horizon 
region of a non-extremal and non-rotating black hole. Though 
we work in four dimensions, our approach is generalizable 
to arbitrary dimensions. The rotating case will be addressed 
separately in a future publication.  

We will consider a class of spacetimes whose near-horizon 
geometry resembles that of a Schwarzschild string. We then
consider spacetime diffeomorphisms which preserve the
boundary conditions at the horizon. The key assumption is
that the near-horizon symmetries which give rise to non-zero
`charges' will be realized as symmetries also in the quantum
theory. The quantum numbers of these charges would then label 
the quantum states corresponding to the classical spacetime. 
It is in this spirit that one may count the degeneracy of a 
subset of states that are associated with a black hole of a 
given mass. A similar criterion was 
advocated in \cite{BH} for asymptotic symmetries. 
The canonical algebra of charges can be realized (in the physical 
phase space of general relativity) as a Virasoro algebra with a 
central extension. The entropy of the string is then given by the 
logarithm of the degeneracy of the representative states of the  
Virasoro algebra that share a common (large) conformal weight. 
A large conformal weight is equivalent to a large mass or area 
of the black string.

The line-element  of a Schwarzschild-string in `tortoise'-coordinates is 
\beq
ds^2=\Delta(-dt^2+dr_*^2)+r^2d\theta^2+r^2\sin^2\theta d\phi^2+
\ell^2d\chi^2\;,\quad\Delta={dr\over dr_*}=1-{2GM\over r}\;.
\label{sch}
\eeq
Here $0\le\chi<1$ and thus $\ell$ is the length of the string.
$M$ is the mass per unit length of the string. For the moment
we keep $\ell$ arbitrary. We will consider the class of metrics which 
approach the Schwarzschild-string geometry near the horizon of
(\ref{sch}). Our boundary conditions are
motivated by the requirement of conformal symmetry: 
\beq
&&\delta g_{tt}={\cal O}(\Delta^2)\;,\;
\delta g_{tr_*}={\cal O}(\Delta)\;,\;
\delta g_{t\theta}={\cal O}(\Delta^2)\;,\;
\delta g_{t\phi}={\cal O}(\Delta)\;,\;
\delta g_{t\chi}={\cal O}(\Delta)\;,\nn
&&\delta g_{r_*r_*}={\cal O}(\Delta)\;,\;
\delta g_{r_*\theta}={\cal O}(\Delta)\;,\;
\delta g_{r_*\phi}={\cal O}(\Delta)\;,\;
\delta g_{r_*\chi}={\cal O}(\Delta)\;,\;
\delta g_{\theta\theta}={\cal O}(\Delta^2)\;,\nn
&&\delta g_{\theta\phi}={\cal O}(\Delta)\;,\;
\delta g_{\theta\chi}={\cal O}(\Delta)\;,\;
\delta g_{\phi\phi}={\cal O}(\Delta)\;,\;
\delta g_{\phi\chi}={\cal O}(\Delta)\;,\;
\delta g_{\chi\chi}={\cal O}(\Delta)\;.
\label{bc}
\eeq
Now we seek vector fields $\xi^\mu$ which generate 
diffeomorphisms preserving these fall-off conditions. 
Let us make a near-horizon expansion of such a
$\xi^\mu(t,r_*,\theta,\phi,\chi)$, which is a candidate 
for a near-horizon symmetry vector, in powers of $\Delta$:
\beq
&&\!\!\!\xi^t=T(t,\theta,\phi,\chi)+{\cal O}(\Delta)\;,\;
\xi^{r_*}=R(t,\theta,\phi,\chi)+{\cal O}(\Delta)\;,\;
\xi^\chi=X(t,\theta,\phi,\chi)+{\cal O}(\Delta)\;,\nn
&&\!\!\!\xi^\theta=\Theta(t,\theta,\phi,\chi)+\Theta_1(t,
\theta,\phi,\chi)\Delta+{\cal O}(\Delta^2)\;,\;
\xi^\phi=\Phi(t,\theta,\phi,\chi)+{\cal O}(\Delta)\;.
\label{kill}\eeq
Expansion coefficients which are associated with non-vanishing
on-shell\thanks{`On-shell' refers to the implementation of the 
energy and momentum constraints. This phrase is to be distinguished 
from `on-the-solution', i.e., `on (\ref{sch})', which will also be
used here. Eventually all charges will be evaluated on-the-solution.}
surface charges (see below) at the horizon 
are physical and measurable, and those which give zero surface 
charges are irrelevant. The expansions (\ref{kill}) are motivated
by the desire to get an infinite symmetry with the above 
requirement. 

Equating $\delta g_{\mu\nu}={\cal L}_\xi g_{\mu\nu}$ we get 
relations between various components of $\xi^\mu$, 
\beq
&&tt:\;R=-4GM\partial_t T\;;\quad
t\theta:\;\partial_t\Theta=0\;,\;\partial_\theta 
T=4GM^2\partial_t\Theta_1\;;\quad
t\phi:\;\partial_t\Phi=0\;;\nn
&&t\chi:\;\partial_tX=0\;;\quad
\theta\theta:\;\partial_\theta\Theta=0\;,\;R=-2GM
\partial_\theta\Theta_1\;;\quad
\theta\phi:\;\partial_\phi
\Theta+\sin^2\theta\partial_\theta\Phi=0\;;\nn
&&\theta\chi:\;\partial_\theta
X=\partial_\chi\Theta=0\;;\quad
\phi\phi:\;\partial_\phi
\Phi+\cot\theta\Theta=0\;;\quad
\phi\chi:\;\partial_\phi
X=\partial_\chi\Phi=0\;;\nn
&&\chi\chi:\;\partial_\chi X=0\;,
\label{eq}
\eeq
which can 
be solved completely. Solutions for $\Theta,\Phi,X$ and 
the zero mode of $\xi^t$ are:
\beq
\xi^t=T_0,\;\Theta=A\cos\phi+B\sin\phi,\;\Phi=\cot\theta
(-A\sin\phi+B\cos\phi)+k_1,\;X=k_2\;.
\eeq
$T_0,A,B,k_1,k_2$ are constants. The following linear combinations of 
these form the five global Killing vectors of the Schwarzschild
string (one associated with its mass per unit length, three associated 
with the rotational symmetry and one more with the translational 
symmetry along the length of the string): $\partial/\partial t,\;
\partial/\partial\phi,\;(\cos\phi\,\partial/\partial\theta-\cot
\theta\sin\phi\,\partial/\partial\phi),\;(\sin\phi\,\partial/
\partial\theta+\cot\theta\cos\phi\,\partial/\partial\phi),\;\partial/
\partial\chi,\;$. 
For us, however, the interesting solutions are the higher order modes 
which are near-horizon symmetry vectors. It is convenient to express  
these near-horizon symmetries in the Fourier-modes of $\theta,\;\phi$
and $\chi$. If we define $x_\pm=t/\sqrt{8G^2M^2}\pm\theta$, one set 
of modes is
\beq
&&T^{nn'n''}=\sqrt{2G^2M^2}\exp\left[2inx_-+in'\phi+2\pi in''\chi
\right]\;,\nn
&&R^{nn'n''}=-4GM\,in\exp\left[2inx_-+in'\phi+2\pi in''\chi\right]\;,\nn
&&\Theta_1^{nn'n''}=-\exp\left[2inx_-+in'\phi+2\pi in''\chi\right]\;.
\label{sol}\eeq
There is another set involving $x_+$, which is obtained by replacing 
$x_-$ by $x_+$. Its contribution will be incorporated later. The overall
normalization of the solutions (\ref{sol}) is fixed by the 
surface-deformation (SD) algebra to be defined below. The $\phi$-modes
$n'$ play no interesting r\^{o}le in the analysis to follow.
Henceforth, we set them to zero. On the other hand, we pick up the
diagonal elements $(n=n'')$ of $x_-$ and $\chi$-modes in which 
$\xi^\mu$ furnish a Diff$(S^1)$ algebra. If we define the surface 
deformation parameters $\hat\xi^t_n=\sqrt\Delta
\xi^t_{n0n}$ and $\hat\xi^i_n=\xi^i_{n0n}$, the SD-brackets 
\cite{BH} of $\hat\xi^\mu_n$ give rise to the corresponding
brackets for $\xi^\mu$ in the following way: 
\beq
&&\Big\{\hat\xi_n,\hat\xi_m\Big\}^t_{\rm SD}=\hat\xi^i_n\partial_i
\hat\xi^t_m-(m\leftrightarrow n)\;,\quad
\Big\{\xi_n,\xi_m\Big\}^t_{\rm SD}=i(m-n)\xi^t_{m+n}\;,\nn
&&\Big\{\hat\xi_n,\hat\xi_m\Big\}^i_{\rm SD}=\hat\xi^j_n\partial_j
\hat\xi^i_m+h^{ij}\hat\xi^t_n\partial_j\hat\xi^t_m-
(m\leftrightarrow n)\;,\quad
\Big\{\xi_n,\xi_m\Big\}^i_{\rm SD}={\cal O}(\Delta)
\label{sd}\eeq
where $\{\hat\xi_n,\hat\xi_m\}^t_{\rm SD}=\sqrt\Delta\{\xi_n,\xi_m
\}^t_{\rm SD}$ and $\{\hat\xi_n,\hat\xi_m\}^i_{\rm SD}=\{\xi_n,\xi_m
\}^i_{\rm SD}$.

To realize these local symmetries in terms of the canonical
Poisson brackets, let us recall the canonical surface 
deformation generators of the ADM-formulation \cite{Regge}
(the phase space coordinates are $(h_{ij},\pi^{ij})$):
\beq
H[\hat\xi]={1\over 16\pi\ell G}\int d^{\,4}x\,\hat\xi^\mu(x){\cal 
C}_\mu(x)+Q[\hat\xi].
\eeq
Here ${\cal C}_\mu= ({\cal C},C_i)$ are the energy 
and momentum constraints ${\cal C}=
(\pi_{ij}\pi^{ij}-{1\over 3}\pi^2)/\sqrt h-{}^4\!R\sqrt h$ and $C_i=-2
\pi^j_{i|j}$, where ${}^4\!R$ is the curvature of the $t=$
constant surface and $\pi=\pi^i_i$. A vertical bar denotes 
covariant differentiation with respect to the induced metric on the space
slice. $Q[\hat\xi]$ represents the appropriate boundary term which, 
in the presence of the boundary conditions (\ref{bc}), makes the total 
generator $H[\hat\xi]$ differentiable at the boundary. In other 
words, boundary terms appearing due to the variation of the 
constraints 
in $H[\hat\xi]$ in the phase space coordinates and from variations 
of $Q[\hat\xi]$ cancel each other in the presence of the boundary conditions 
(\ref{bc}). 
The variation of the constraints alone thus gives the variation 
of $\delta Q[\hat\xi]$ as a combination of total derivatives:
\beq
\delta Q[\hat\xi]={1\over 16\pi\ell G}\int d^{\,4}x\bigg\{G^{ijkl}
\Big[\,\hat\xi^t(\delta h_{ij})_{|k}-\delta h_{ij}\hat\xi^t_{,\,k}
\Big]+2\hat\xi^i\delta\pi^l_i-\hat\xi^l\pi^{ik}\delta h_{ik}\bigg
\}_{|\,l}
\label{var}\eeq 
where $2G^{ijkl}=\sqrt h(h^{ik}h^{jl}+h^{il}h^{jk}-2h^{ij}h^{kl})$ and
the integral is evaluated at the horizon of (\ref{sch}). 
The variations $(\delta h_{ij},\delta\pi^{ij})$ in (\ref{var}) 
belong to the constraint surface in the phase space while 
the coefficients take their values on the solution. 
The coefficient of the last
term vanishes identically on-the-solution. The rest of the variations
can be integrated, giving rise to the surface charge
\beq
Q[\hat\xi]={1\over 16\pi\ell G}\int d^{\,4}x\bigg\{\bar 
G^{ijkl}\Big[\,\hat\xi^t
h_{ij|k}-h_{ij}\hat\xi^t_{,\,k}\Big]+2\hat\xi^i
\pi^l_i\bigg\}_{|\,l}\label{fullq}
\eeq 
where all barred quantities refer to the on-the-solution metric.
The surface charge (\ref{fullq}) simplifies on-the-solution to  
\beq
Q[\hat\xi]={1\over 16\pi\ell G}\int_\Delta d\theta d\phi d\chi\;
\bigg\{\bar G^{ijkr_*}\Big[-\bar h_{ij}\partial_k\hat\xi^t\Big]
\bigg\}\;.
\label{js}\eeq 
Since the surface charges (\ref{js}) are linear functionals of 
$\xi^\mu$, they obey the canonical Poisson bracket algebra induced 
by the SD-algebra (\ref{sd}) in the constrained phase space
$\{Q[\hat\xi],Q[\hat\eta]\}=Q[\{\hat\xi,\hat\eta\}_{\rm SD}]+W_{
\xi\eta}$ where $W$ is a possible central extension. The canonical 
Poisson bracket can be cast into the form of a Lie-bracket 
\beq
\Big\{Q[\hat\xi_n],Q[\hat\xi_m]\Big\}=Q\Big[\{\hat\xi_n,\hat\xi_m
\}_{\rm SD}\Big]+W_{nm}={\cal L}_{\hat\xi_m}Q[\hat\xi_n]\label{lp},
\eeq
where ${\cal L}_{\hat\xi_n}\hat\xi^t_m=\hat\xi^i_n\partial_i\hat
\xi^t_m$. Eq. (\ref{lp}) can be understood as follows. In phase space 
one can associate a vector field $q^a[h_{ij},\pi^{ij}]$ with each 
phase space scalar functional, say the charge $Q$, such that ${\cal L}_qF=
\{Q,F\}$, for an arbitrary scalar functional $F$. Since $Q$ is a
linear functional of the phase space coordinates (see (\ref{fullq})) 
and $\hat\xi^\mu$, the Lie-derivative generated by the vector field 
$q^a$ is taken to be equal to ${\cal L}_{\hat\xi}$. 

The (infinitely many)  canonical generators are obtained by 
using the various modes of $\hat\xi^\mu$ in (\ref{js}):
\beq
L_n=Q[\hat\xi_n]={1\over 8\pi\ell G}\int_\Delta d\theta d\phi 
d\chi\,\left(\sqrt hh^{r_*r_*}\partial_{r_*}\hat\xi^t_n\right)
={A_H\sqrt{\;2}\over 32\pi G}\,\delta_{n0},
\label{defQ}\eeq
where $A_H=16\pi(GM)^2$. The central extension $W_{nm}$ is 
evaluated from the formula (\ref{lp}) and the SD-bracket
(\ref{sd}):
\beq
W_{nm}=i(n-m)L_{m+n}+{\cal L}_{\hat\xi_m}Q[\hat\xi_n]\;.
\eeq
In quantum theory we replace the Poisson brackets by 
commutator brackets: $i\{\;...\;\}\to[\;...\;]$. As a result
the central charge takes the following form:
\beq
iW_{nm}=i{\cal L}_{\hat\xi_m}Q[\hat\xi_n]-(n-m)L_{n+m}\;
{\mathop=^{\rm def}}\;{c\over 12}(n^3-n)\delta_{m+n}\;.
\label{vir}\eeq
Thus the value of $c$ depends on the on-the-solution value of 
$L_{n+m}$ and the Lie bracket 
\beq
{\cal L}_{\hat\xi_m}Q[\hat\xi_n]=
{1\over 8\pi\ell G}\int dr_*d\theta d\phi d\chi \sqrt hh^{kl}
\hat\xi_m^i\nabla_i\nabla_l\partial_k\hat\xi^t_n\;.
\label{cform}\eeq
One comment is in order regarding the formula (\ref{cform}). The $r_*$ 
integral is evaluated at $r=2GM$ (one also uses $dr=\Delta dr_*$). 
The required anti-symmetry of $W_{nm}$ under the exchange of 
$(m\leftrightarrow n)$ results naturally from the on-the-solution
value of $L_{n+m}$ in (\ref{defQ}) and the integral (\ref{cform}) 
which turns out to have the general form $(\gamma n^3+\beta n)
\delta_{m+n}$, where $\gamma=(\alpha^2-1)A_H\sqrt 2/4\pi G$ and
$\beta=A_H\sqrt 2/32\pi G$ with $\alpha=2\pi GM/\ell$. Note
that a linear term in $n$ appears also from $L_{n+m}$. When 
one adds the two linear terms 
the $(n^3-n)$ form of the central extension of the 
Virasoro algebra (\ref{vir}) is reproduced for the
choice $\alpha^2=9/8$. The value of $c$ is then read off from the 
coefficient $\gamma$:
\beq
c=12\gamma={3A_H\sqrt{\;2}\over 8\pi G}\;.\label{c}
\eeq
When the two length scales of the string, $\ell$ and $GM$, 
are proportional, one should check its stability as pointed 
out in \cite{Greg} (see also \cite{Mart} for an entropy argument). 
Our choice $\alpha^2=9/8$ leads to $3\ell=4\pi GM\!\sqrt{\,2}$ which is 
inside the stability bound $0<\ell<3.375\pi GM$ (estimated 
from the entropy bound $S_{\rm string}>S($5-d Schwarzschild of
mass $M)$. 

Using (\ref{c}) and (\ref{defQ}) the statistical entropy $S_-$ 
from the $(x_-)$-sector is found to be given by (the asymptotic
formula is valid for $A_H\gg G$) \cite{Cardy}
\beq
S_-=2\pi\sqrt{cL_0\over 6}={A_H\over 8G}\;.
\eeq
The two sectors, $x_\pm$, thus give the total entropy 
$S=S_-+S_+=A_H/4G$ which agrees with the Bekenstein-Hawking entropy. 


We now comment on some aspects of the calculation. 1) We chose 
a specific set of boundary conditions near the horizon.
The choice was motivated by several works \cite{Strom,Carlip,
Tooft} where it was shown that conformal
symmetry plays a key r\^{o}le in describing near-horizon states. Our
primary aim has been to obtain the conformal symmetry. 2) Obviously
our boundary condition (\ref{bc}) is only one of possible choices,
but it is this one that gives a conformal symmetry with non-vanishing
central charge. Presumably this provides the maximum degeneracy.
3) The counting of black hole states is 
reproduced by the counting of black string states in the following 
sense -- the Boltzmann entropies of the black string and the black hole 
should be related as $\log\Omega_{\rm string}\sim\log\Omega_{\rm
hole}+N$ where $N$ is the number of microscopic constituents 
along the string which is proportional to its length $\ell$
which in turn is proportional to $M$. 
Thus the statistical entropy of the black 
hole is $S\sim\log\Omega_{\rm string}-\ell\sim M^2-M \sim M^2$ for 
large $M$, which matches the Bekenstein-Hawking entropy of the black
hole. 4) We required a large string length in the above calculation,
implying a five dimensional Planck 
length $(\ell_P\ell)^{1/2}$ which is large compared to the four 
dimensional Planck length $\ell_P$. This signals the opening up 
of an extra dimension near the horizon \cite{Extrad}
(what happens asymptotically is still an open question).
5) An inappropriate choice of coordinates may lead to 
some difficulties, because one needs an expansion parameter near the 
horizon ($\Delta$) that is unaffected under derivatives. 
Otherwise, different orders in $\Delta$ get mixed up in the Killing 
equations and one encounters more unknown functions 
than the number of available equations. 
6) The conformal symmetry lives in the plane $(t/\sqrt{8G^2M^2}+2\pi
\chi)\pm\theta$. One may be tempted to find a connection between this
plane and the string world sheet. Note however, that the $(t-r_*)$-plane 
plays no special r\^{o}le in this approach and both the sectors of 
conformal symmetry contribute equally to the entropy, unlike in
\cite{Carlip}.  
\vspace{-.2cm}
\subsection*{Acknowledgments}
SD and AG thank Abhay Ashtekar for discussions and various
suggestions. The works of
SD and AG were supported by the National Science Foundation 
grant PHY95-14240 and the Eberly Research Funds of Penn State.
PM thanks the Theory Division of CERN, where the collaboration started. 

\end{document}